\documentclass[doublecol]{epl2}

\usepackage{subfigure}
\usepackage{graphicx}
\usepackage{subfigure}

\title{Adaptive information filtering for dynamic recommender systems}
\shorttitle{Adaptive information filtering for dynamic networks} 

\author{Ci-Hang JIN\inst{1} \and Jian-Guo LIU\inst{2} \and Yi-Cheng ZHANG \and Tao ZHOU\inst{1,3}\thanks{zhutou@ustc.edu}}
\shortauthor{Ci-Hang JIN \etal}

\institute{
  \inst{1} Physics Department, University of Fribourg, Chemin du Mus\'ee 3, 1700 Fribourg, Switzerland\\
  \inst{2} Research Center of Complex Systems Science, University of
  Shanghai for Science and Technology, Shanghai 200093, P. R.
  China\\
  \inst{3} Department of Modern Physics and Nonlinear Science
  Center, University of Science and Technology of China, Hefei
  230026, P. R. China
}


\abstract{The dynamic environment in the real world calls for the
adaptive techniques for information filtering, namely to provide
real-time responses to the changes of system data. Where many
incremental algorithms are designed for this purpose, they are
usually challenged by the worse and worse performance resulted from
the cumulative errors over time. In this Letter, we propose two
incremental diffusion-based algorithms for the personalized
recommendations, which integrate some pieces of local and fast
updatings to achieve the approximate results. In addition to the fast
responses, the errors of the proposed algorithms do not cumulate
over time, that is to say, the global recomputing is unnecessary.
This remarkable advantage is demonstrated by several metrics on
algorithmic accuracy for two movie recommender systems and a social
bookmarking system.}

\begin{document}

\maketitle

\section{Introduction}
The information overloading problem in the Internet era causes many
automatic filtering techniques to appear \cite{b.Hanani01}, among
which the personalized recommender system is a promising one
\cite{b.Pazzani99, b.Goldberg92, b.Resnick97}. In the literature,
the central problem concerned in the design of recommender systems
is how to improve the accuracy \cite{b.Konstan04, Adomavicius2005,
Bennett2007} and diversity \cite{Diversity1, Diversity2, Diversity3}
of recommendations. Recently, the well-developed Web 2.0 technique
facilitates users to frequently communicate with the system in an
easy way, leading to a real-time data flow. How to
properly design an incremental algorithm that can adaptively give
responses to the changes of data has thus become an urgent problem
nowadays. Where some previous studies have already considered such an
issue \cite{b.Catarina08, b.Rickard05, b.Thomas05, Suryavanshi05,
b.Badrul05, b.Hu08, b.AlSumait08}, they are challenged by the worse
and worse performance resulted from the cumulative errors over time.

Recently, Zhang \emph{et al.} \cite{b.Zhang07, b.Zhang072} have
successfully applied the classical physical processes, such as the
heat conduction and mass diffusion, to deal with the personalized
recommendation problem. The original algorithms require a kind of
steady states, and to arrive at these states is time consuming. Zhou
\emph{et al.} \cite{Diversity3,b.Zhou07} thus proposed the
simplified versions where only one step of heat conduction and/or
mass diffusion is taken into account. These simplified algorithms are
considerably more accurate than the standard collaborative filtering
\cite{Breese1998} and much faster yet with competitive accuracy
compared with the matrix decomposition techniques \cite{Ren2008}. In
this Letter, we propose two adaptive algorithms based on the
one-step diffusion methods, which integrate some pieces of local
updatings and can provide very fast response to each unit change.
More significantly, the error between the results of the adaptive
algorithms and the static algorithms does not grow larger over time,
namely the adaptive algorithms could keep up with the changes
tightly and thus no global recomputing is needed. This remarkable
feature highlights a new methodology to handle the huge-size dynamic
systems, and can avoid the serious problems in the
maintaining-retraining scheme \cite{Suryavanshi05}, like the data
synchronization.

\section{Static algorithms}
A recommender system can be represented by a bipartite network
\cite{Shang2009}, which consists of a user set $\vect{U}$ and an
item set $\vect{I}$. To make the description clear, we use lower
case letters $i$, $j$, $k$, $\dots$ to index the users and Greek
letters $\alpha$, $\beta$, $\gamma$, $\dots$ to index the items.
Each user is connected to the items she/he collected, reviewed,
voted with high ratings, purchased, etc. Later, we simply use
collect/collected to stand for all possible relations between users
and items. The bipartite network can be described by an adjacency
matrix $\vect{A}=\left\{ a_{i\alpha} \right\}$, where $a_{i\alpha} =
1$ if an edge between user $i$ and item $\alpha$ exists,
$a_{i\alpha} = 0$ otherwise.

A diffusion-based recommendation algorithm applies some physical
processes to simulate the resource propagation in the bipartite
network. For an arbitrary target user, assigning some resources to
her/his collected items, whether she/he user will collect certain
items in the future can be predicted according to the resource
distribution after the diffusion process \cite{b.Zhou07}. Generally
speaking, this kind of methods can be formulated as $\vect{f'} =
\vect{W} \vect{f}$, where $\vect{f}$ is the vector representing the
initial resource allocation, $\vect{f'}$ is the final resource
distribution and $\vect{W}$ is the propagation matrix. Provided the
target user $i$, a straightforward way, also adopted in this Letter,
is to set the initial vector as $f_\alpha=a_{i\alpha}$, namely to
assign all the objects collected by user $i$ a unit of
resource\footnote{The initial resource allocation can be designed in
more sophisticated ways (see for example the Ref.
\cite{Diversity1}), but as shown in the later, the initial
conditions have no business with the adaptive algorithms, and thus
we do not intend to discuss it in this Letter.}. The element
$w_{\alpha \beta}$ denotes the fraction of item $\beta$'s initial
resource that eventually arrives at item $\alpha$. Sorting all
uncollected items by $\vect{f'}$ in descending order, the ones at
the top positions are recommended.

We here consider two simplest diffusion-based algorithms: the
one-step \emph{mass diffusion} (MD) \cite{b.Zhou07} and \emph{heat
conduction} (HC) \cite{Diversity3}\footnote{In Ref.
\cite{Diversity3}, these two algorithms are referred as ProbS and
HeatS. Since we here deal with the bipartite networks, the diffusion
from items to items actually contains two steps.}. The propagation
matrix of MD is denoted by $\vect{M}$ with
\begin{equation}
\label{eq.1} m_{\alpha \beta} = \frac{1}{k_{\beta}} \sum_{i \in
\vect{U}} \frac{a_{i\alpha}a_{i\beta}}{k_{i}},
\end{equation}
where $k_{i}$ and $k_{\beta}$ are the degrees of user $i$ and item
$\beta$, respectively. Analogously, the propagation matrix
$\vect{H}$ of HC reads
\begin{equation}
\label{eq.2} h_{\alpha \beta} = \frac{1}{k_{\alpha}} \sum_{i \in
\vect{U}} \frac{a_{i\alpha}a_{i\beta}}{k_{i}}.
\end{equation}

\begin{figure}[!h]
\centering
\includegraphics[width=8.5cm]{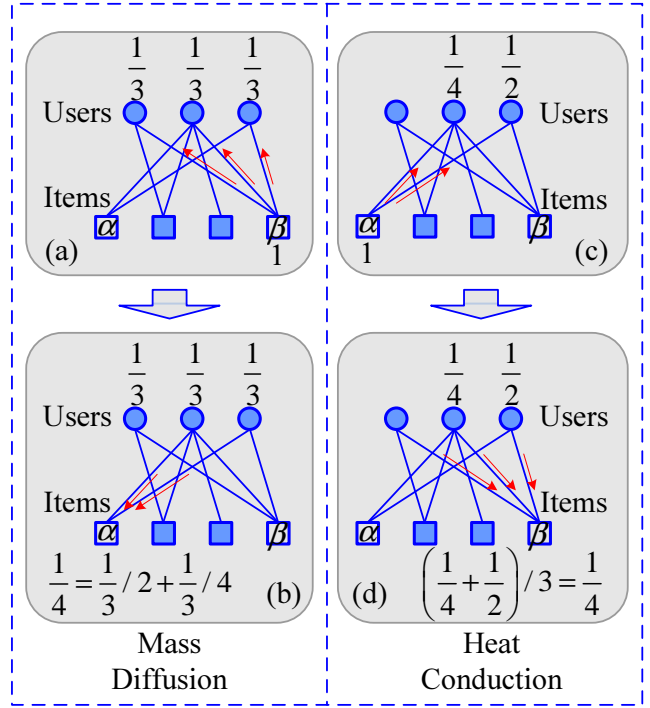}
\caption{(Color online) Illustration of the diffusion processes by
MD and HC. (a) and (b) display how a unit of resource initially
located on item $\beta$ spreads to item $\alpha$ by MD, while (c)
and (d) show how a unit of resource (a physical explanation of this
kind of resource is ``temperature", see Refs. \cite{Diversity3,
b.Zhang07} for details) initially located on item $\alpha$ spreads
to item $\beta$ according to HC. As shown in this figure,
$m_{\alpha\beta}=h_{\beta\alpha}=1/4$.} \label{fig.MC_HC}
\end{figure}

Figure \ref{fig.MC_HC} illustrates the processes of MD and HC.
Interestingly, the MD and HC are a pair of reverse
processes, say $\vect{M}^{T}=\vect{H}$. A direct benefit of this
feature is that one only needs to compute one propagation matrix and
then immediately gets the other one by transposition, which can
reduce the computation when considering the hybrid algorithm
involving both the two processes \cite{Diversity3}. In this Letter,
we will exploit this feature in a more artful way to develop an
effective and efficient adaptive algorithm.

\section{Exact solution on incremental data}
Although the ability to provide accuracy and diverse recommendations
of the MD and HC, especially the hybrid version combined these two,
has been demonstrated by extensively experimental analysis
\cite{Diversity3}, they are not able to work directly in a dynamic
environment where the bipartite network (represented by $\vect{A}$)
itself varies moment to moment. A unit change here is defined as the
addition of a new edge, whatever it is attached to a new user, a
new item, or old ones. Accordingly, the total number of edges, $l$,
is naturally a good indication of the state of the evolving system,
which is added to all relevant symbols as the superscript if
necessary. For example, $k_{\alpha}^{\left( l \right)}$ means the
degree of the item $\alpha$ when there are $l$ edges in the network.
A cautious yet infeasible method is to recompute every time when a
new edge is added to the system. Instead, the purpose of an adaptive
algorithm (or say an incremental algorithm) is to derive the
recommendation vector $\vect{f'}^{(l+1)}$ in a local and fast way
that can avoid the global recomputing\footnote{According to the time
resolution of the system, it is very possible that several edges are
added at the same time. In such a case, we always assume that these
edges are added one by one in a random order. In addition, a unit
change may not be the addition of an edge, but a removal of an edge
like in \emph{del.icio.us} users are free to remove their
collections. As shown later, our algorithms can handle the removal
of edges in the similar way to the addition of edges, and thus we
here only discuss how to get $\vect{f'}^{(l+1)}$.}.

Before the introduction of the approximately adaptive algorithm, we
first figure out the exact solution of $\vect{f'}^{(l+1)}$ for an
arbitrary target user. As stated above, the propagation matrices of
MD and HC can be easily derived by each other, so we only present
the case for MD, while the one for HC can be analogously obtained
without any difficulties. Since for any target user,
$\vect{f'}^{(l+1)}=\vect{M}^{(l+1)}\vect{f}^{(l+1)}$, and the change
of $\vect{f}$ is straightforward, we only focus on the change of
$\vect{M}$.

\begin{figure}[!h]
\centering
\includegraphics[width=8cm]{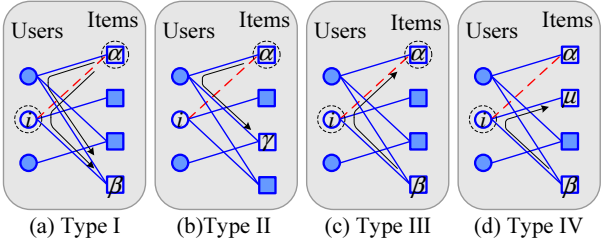}
\caption{(Color online) Illustration of four types of changes to the
$\vect{M}$ with the addition of a new edge. The dotted line is the
new edge added between user $i$ and item $\alpha$. The lines with
arrow are the diffusion paths affected by the new edge. And the
dotted circles emphasize the nodes (users or items) being
responsible for the corresponding type of changes. }
\label{fig.dynamic}
\end{figure}

With the addition of a new edge between user $i$ and item
$\alpha$ (we first consider the case that both $i$ and $\alpha$ already exist in the
system), the propagation matrix $\vect{M}^{(l)}$ changes to be
$\vect{M}^{(l+1)}$. In the microscopic level, the elements of
$\vect{M}^{(l+1)}$ can be classified into four types according to
the way how they change. As illustrated in Fig. 2, the four types
are: Type I---$m_{\beta \alpha}^{\left( l+1 \right)}$ where item
$\beta$ is connected to user $i$ ($\beta \neq \alpha$). Type
II---$m_{\gamma \alpha}^{\left( l+1 \right)}$ where item $\alpha$
and $\gamma$ are commonly collected by one or more users rather than
the user $i$. Type III---$m_{\alpha \beta}^{\left( l+1 \right)}$
where item $\beta$ is connected to user $i$ ($\beta \neq
\alpha$). Type IV---$m_{\mu \beta}^{\left( l+1 \right)}$ where the
item $\beta$ and $\mu$ are both connected to the user $i$ ($\beta
\neq \alpha$, $\mu \neq \alpha$).

Denoting by $\delta_{\beta \alpha}^{\left( l+1 \right)}$ the change
of the element $m_{\beta \alpha}$, say $\delta_{\beta
\alpha}^{\left( l+1 \right)} = m_{\beta \alpha}^{\left( l+1 \right)}
- m_{\beta \alpha}^{\left( l \right)}$, according to Fig. 2, it is
easy to derive the following updating rules:
\begin{equation}
\label{eq.3} \delta_{\beta \alpha}^{\left( l+1 \right)} =
\frac{-1}{k_{\alpha}^{\left( l+1 \right)}}m_{\beta \alpha}^{\left( l
\right)} + \frac{1}{k_{\alpha}^{\left( l+1 \right)} k_{i}^{\left(
l+1 \right)}}.
\end{equation}
\begin{equation}
\label{eq.4} \delta_{\gamma \alpha}^{\left( l+1 \right)} =
\frac{-1}{k_{\alpha}^{\left( l+1 \right)}} m_{\gamma \alpha}^{\left(
l \right)}.
\end{equation}
\begin{equation}
\label{eq.5} \delta_{\alpha \beta}^{\left( l+1 \right)} =
\frac{1}{k_{\beta}^{\left( l+1 \right)} k_{i}^{\left( l+1 \right)}}.
\end{equation}
\begin{equation}
\label{eq.6} \delta_{\mu \beta}^{\left( l+1 \right)} =
\frac{-1}{k_{\beta}^{\left( l+1 \right)} k_{i}^{\left( l \right)}
k_{i}^{\left( l+1 \right)}}.
\end{equation}

\begin{table}
\small \caption{Fundamental statistics of the applied data sets. The
data sets for Netflix and Delicious are random samples.}
\label{tab.1}
\begin{center}
\begin{tabular}{cccc}
\hline Data Set  & $|U|$ & $|I|$ & Average Object Degree
\\\hline
MovieLens & 943 & 1574 & 52.43\\
Netflix & 609 & 8824 & 11.33\\
Delicious & 1623 & 32941 & 3.77\\
\hline
\end{tabular}
\end{center}
\end{table}

Note that, the updating rules shown in Eqs. (3)-(6) do not change when
the addition of the edge introduces a new user of a new item, and
only a few extra manipulations are required. When introducing a new
user $i$, only type II happens. When a new item $\alpha$ is
introduced, a new row and a new column need to be appended to
$\vect{M}^{\left( l \right)}$ and initialized to zero, which
actually are the initial values of $m_{\beta \alpha}^{\left( l+1
\right)}$ and $m_{\alpha \beta}^{\left( l+1 \right)}$. In this case
the type II won't happen but type I, type III and type IV should be
considered to update the expanded propagation matrix. Finally, when
both a new user and a new item are added, no change will be made to
the existed elements in $\vect{M}^{\left( l \right)}$. The only
operation is to append a new row and a column to $\vect{M}^{\left( l
\right)}$ and to initialize them to zero.

We have already numerically checked that the updating rules in Eqs.
(3)-(6) can exactly reproduce the results of static algorithm. It
seems that there is no need to design another approximated algorithm
since all the the changed elements can be handled by Eqs. (3)-(6).
Unfortunately, this \emph{exact incremental algorithm} is facing a
dilemma between space and time\footnote{The exact incremental algorithm,
as described by Eqs. (3)-(6), actually consists of two parts.
Firstly we should identify the elements that need to be updated, and
then to update them. The identification can be further divided into
two steps: we should first to find out the elements in each type
(here we obtain the subscripts, such as we know the element
$m_{xy}$ belongs to Type IV), and then to find out the
locations of these elements in the memory space (for example, we, or
the computer programme, have to know the location of $M_{xy}$
before updating it). According to any known techniques of
compressing storage for sparse matrix, the identification process
takes very long time, usually even much longer than the updating (to
locate a specific element in a compressed stored data structure is
not a trivial task, even if we can insert a kind of order among
these elements, it requires $\mathcal{O}(\texttt{log}S)$ time where
$S$ denotes the number of relevant elements). On the hand, if all
the elements of $M$ and $A$ are stored in two
two-dimensional arrays, the identification process is very quick and
thus the whole updating process can be completed in a reasonable
time. However, it asks for $\mathcal{O}(|I|^2+|I|\times |U|)$
memory, and thus fails to handle the huge-size systems that usually
contain millions to billions of users and/or items.}, and thus far away from the
real-world application. In spite of that, the Eqs. (3)-(6) provide a
valuable information for analyzing the errors of the adaptive
algorithms that will be discussed in the next section.

\section{Adaptive algorithms}
The compressed storage (only the nonzero elements are stored) of the
adjacency matrix makes it difficult to be randomly accessed,
however, calculating the propagation from a certain item to the
others, whether by mass diffusion or heat conduction, is very
efficient in such infrastructure, because it can access the
adjacency matrix (or equivalent data structure representing the
connections between users and items) in a sequential
way\footnote{For example, we can simultaneously save all the
neighboring items of each user and all the neighboring users of each
item by dynamical linked lists, and thus the propagation process
will sequential access the elements in a list.}. Hence, for the
elements belonging to Type I and Type II, instead of identifying and
updating, one can just redo a mass diffusion starting with the item
$\alpha$ to get the updated values. We call this algorithm the
\emph{adaptive algorithm with first-order approximation},
abbreviated as AAF.

\begin{figure}[!h]
\centering
\includegraphics[width=8.5cm]{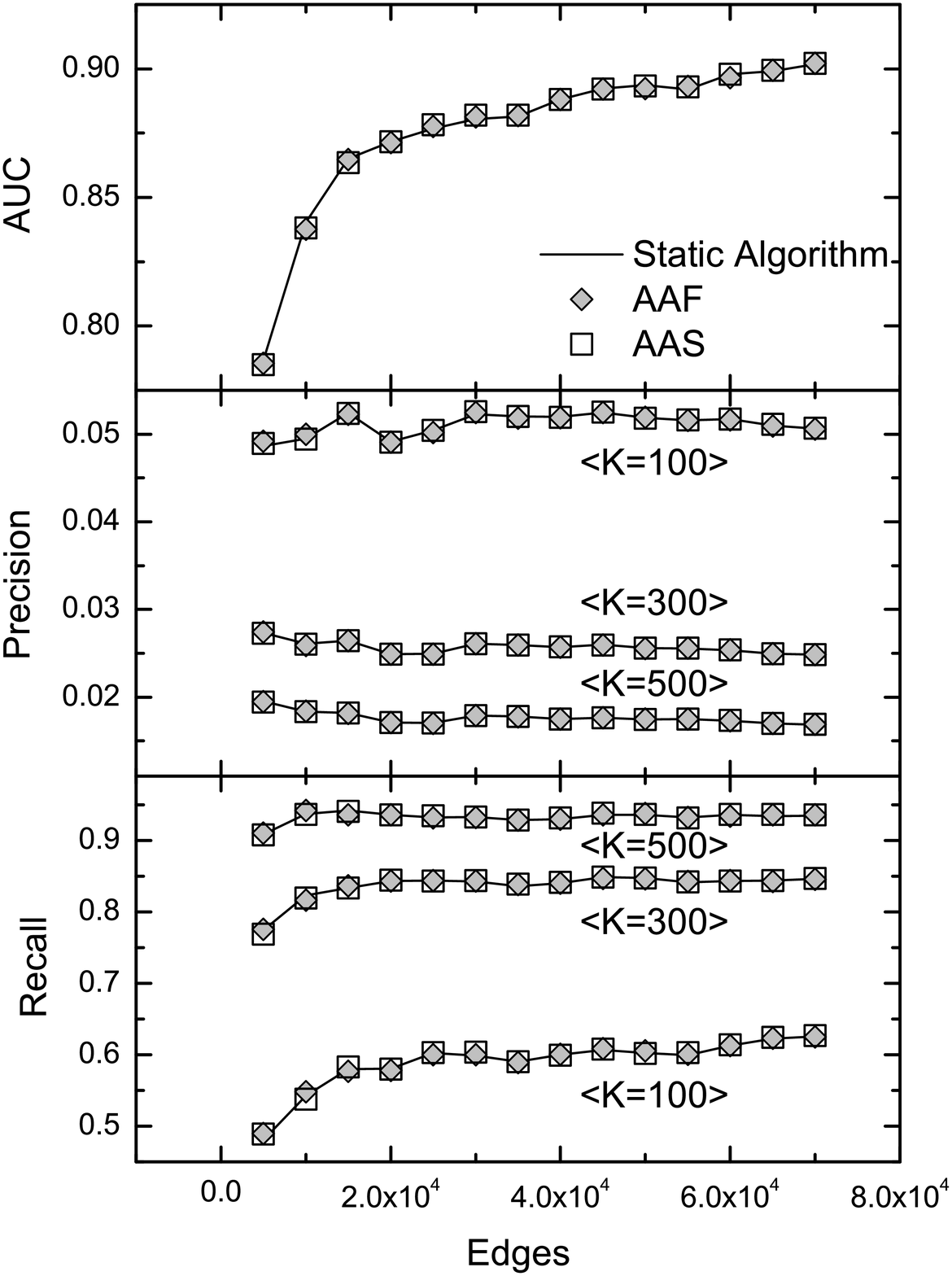}
\caption{AUC, Precision and Recall as the function of the number of
edges fed to the algorithms on MovieLens data. The algorithm starts
with 5000 edges. Precision and Recall are evaluated with $K=100$,
$300$ and $500$, respectively. For all numerical results, including the ones shown in Fig. 4 and Fig. 5, the solid lines represent the results of static algorithm, and the diamonds and squares correspond to AAF and AAS.} \label{fig.1}
\end{figure}

In AAF, when an edge is added, only the elements belonging to the
first and the second types are updated by using the MD starting from
the item $\alpha$. The high efficiency is guaranteed, whereas the
error is introduced for the neglect of other changes of Type III and
Type IV. Detailedly, an error of $\left(k_{\beta}^{\left( l+1
\right)} k_{i}^{\left( l+1 \right)} \right )^{-1}$ (see Eq. (5))
occurs when an element in the third type is not updated, and an error
of $\left ( k_{\beta}^{\left( l+1 \right)} k_{i}^{\left( l \right)}
k_{i}^{\left( l+1 \right)} \right)^{-1}$ (see Eq. (6)) occurs when
an element in the fourth type is not updated. Obviously, the latter
error is far less than the former one since in a real system, many
users have collected more than a hundred items. Therefore, if the
error of the elements in the third type can be eliminated, the
accuracy of the approximate algorithm will be improved
significantly.

The straightforward way to update the elements in the third type
will add too much computation burden, because it needs to simulate
the MD process from all items sharing at least one common
neighboring user with the item $\alpha$ to the item $\alpha$ (see
Eq. (5)). Thanks to the reversibility found for the MD and HC
processes, we conquer this difficulty skillfully. Instead of MD, an
HC process starting from the item $\alpha$ is calculated in
$\vect{A}^{\left( l+1 \right)}$, then the resource arriving at item
$\beta$ is $m_{\alpha \beta}^{\left( l+1 \right)}$. By this way, the
updating of all the elements in the third type is completed by HC
once from one item, with the same complexity as in AAF. We name this
algorithm as the \emph{adaptive algorithm with second-order
approximation}, abbreviated as AAS. In AAS, when an edge to the item
$\alpha$ is added, an MD from the item $\alpha$ is done to get the
updated elements in the first and the second types, then an HC from
the item $\alpha$ is done to get the updated elements in the third
type.

\begin{figure}[!h]
\centering
\includegraphics[width=8.5cm]{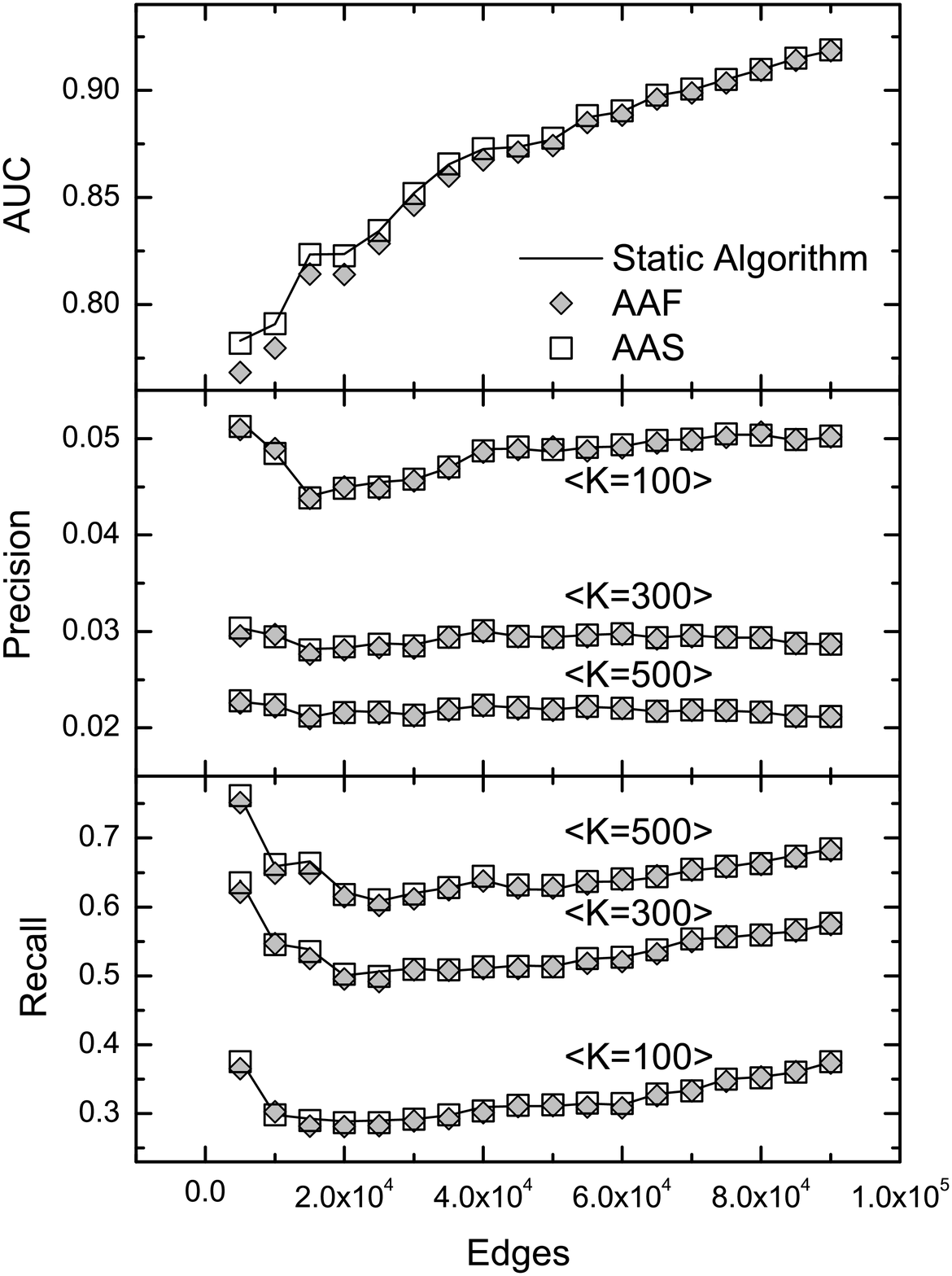}
\caption{AUC, Precision and Recall as the function of the number of
edges fed to the algorithms on Netflix data. The algorithm starts
with 5000 edges. Precision and Recall are evaluated with $K=100$,
$300$ and $500$, respectively.} \label{fig.2}
\end{figure}

\section{Experimental results}
To validate that the proposed adaptive algorithms could provide good
approximation for the static algorithm, we run numerical simulations
on three real systems:
\emph{MovieLens}\footnote{http://www.grouplens.org/.},
\emph{Netflix}\footnote{http://www.netflixprize.com/.}, and
\emph{Delicious}\footnote{http://delicious.com/.}. The characteristics
of these data sets are summarized in Table \ref{tab.1}. The original
data of MovieLens and Netflix record the opinions of users on items
with the ratings from 1 (the worst) to 5 (the best). We convert
these two rating systems to binary systems by defining a connection
between a user and an item only if the corresponding rating is
higher than 2.

A standard sub-sampling validation method is adopted to evaluate the
algorithmic performance, which divides all edges randomly into two
subsets: the training set and the testing set. The training set is
used as the known information for recommending, while the testing
set is used to evaluate the prediction and is invisible to the
recommendation algorithms. In this Letter, $10\%$ of edges
constitute the testing set and the other $90\%$ belong to the
training set. Three widely measurements, \emph{AUC},
\emph{Precision} and \emph{Recall}, are applied to quantify the
accuracy of recommendations (see the review article
\cite{b.Konstan04} for details). All of them are essentially
individual-based indices, that is to say, for any user $i$, the
individual measurements $AUC_{i}^{\left ( l \right )}$,
$Precision_{i}^{\left ( l \right )}$ and $Recall_{i}^{\left ( l
\right )}$ exist. The system-level indices are simply obtained by
averaging the individual measurements over all users. Denoting by
$T_i^{(l)}$ the set of items in the testing set that are collected
by user $i$ (they are invisible from the training set but favored by user
$i$), and $F_i^{(l)}=\vect{I}^{(l)}-\Gamma_i^{(l)}-T_i^{(l)}$ the
set of items not favored yet by user $i$, where $\Gamma_i^{(l)}$ is the
set of items collected by user $i$ in the training set, then
$AUC_i^{(l)}$ is defined as the probability a randomly selected item
in $T_i^{(l)}$ is assigned a topper position in the recommendation
list than a randomly selected item in $F_i^{(l)}$. Clearly, random
recommendations correspond to $AUC\approx 0.5$. Given the length of
recommendation list, $K$, the $Precision_{i}^{\left ( l \right )}$
is defined as the ratio of the number of the recommended items
belonging to $T_{i}^{\left ( l \right )}$ to the total of
recommended items, $K$. The $Recall_{i}^{\left ( l \right )}$ is the
ratio of the number of the recommended items belonging to
$T_{i}^{\left ( l \right )}$ to $|T_{i}^{\left ( l \right )}|$. Note
that, AUC value does not depend on $K$, while the Precision and
Recall are $K$-sensitive. For all these three indices, the larger
value corresponds to the higher recommendation accuracy.

To imitate the dynamic environment in the real world, we sort the
edges in the training set by the time when they appeared, then feed them
to the algorithm one by one. For each 5000 edges, we compute the
AUC, Precision and Recall. The performance along with the adding of
edges is shown in Figs. 3-5. As the addition of edges, the known
information increases and thus the AUC value increases too. In
contrast, the Precision and Recall exhibit larger irregularities
since they concentrate on a tiny fraction of items (i.e., top-recommended
items). For both the three indices, the adaptive algorithms
perform well, even when the Precision and Recall of the static
algorithm displaying irregularities, the approximated algorithms
accurately reproduce these irregularities. In addition, it is seen
that the AAS performs better than the AAF especially for the sparser
data sets, Netflix and Delicious.

\begin{figure}[!h]
\centering
\includegraphics[width=8.5cm]{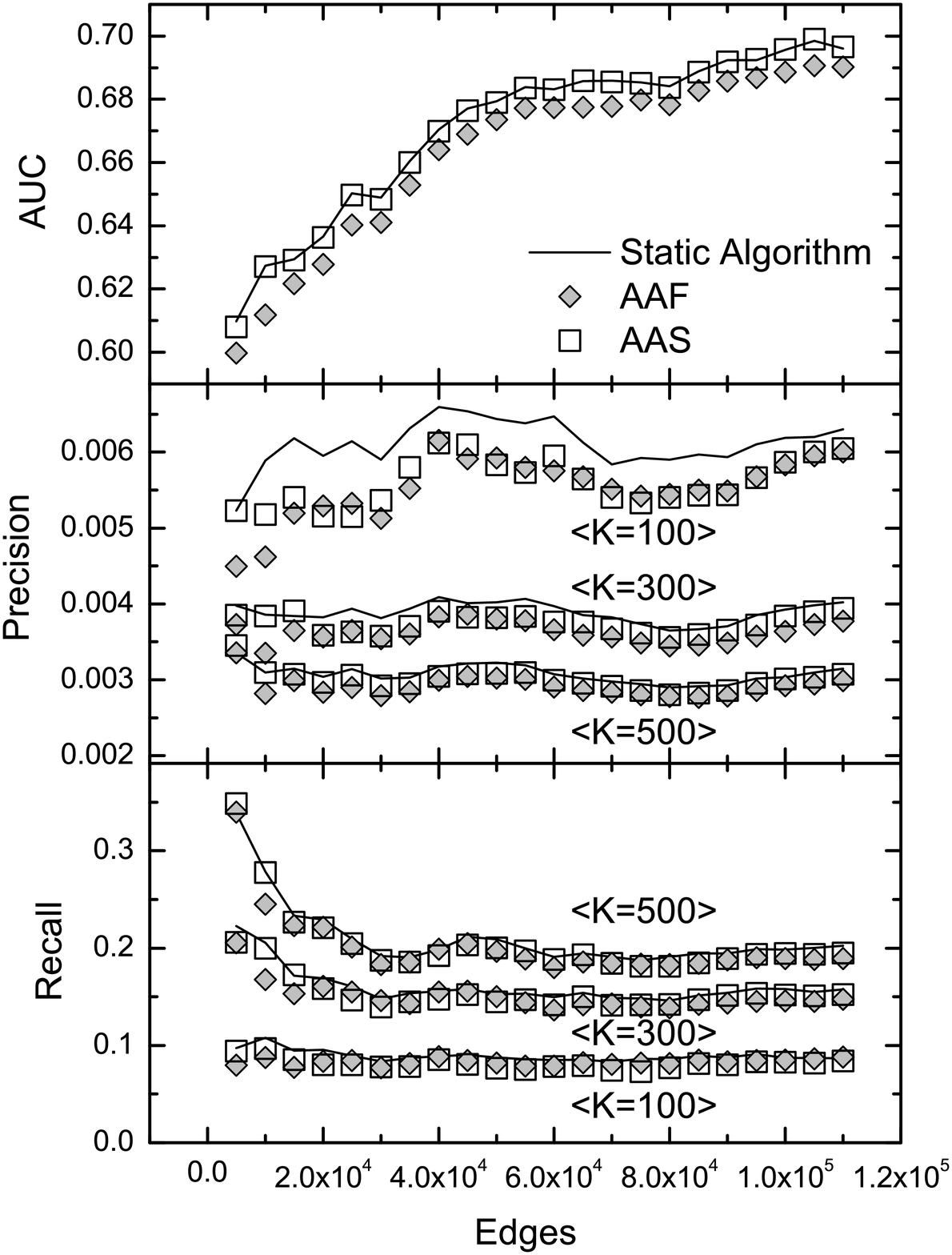}
\caption{AUC, Precision and Recall as the function of the number of
edges fed to the algorithms on Delicious data. The algorithm starts
with 5000 edges. Precision and Recall are evaluated with $K=100$,
$300$ and $500$, respectively.} \label{fig.3}
\end{figure}

The most significant feature of the adaptive algorithms, AAF and
AAS, is that they could follow the static algorithm tightly. For
MovieLens and Netflix data, almost no difference between the adaptive algorithms and the
static algorithm is observed, except a small departure of the AUC for
the Netflix data in the start. Even for the Delicious data where
there are observable errors of the approximations, the difference
does not grow larger over time. This feature strongly encourages the
usage of AAS to completely replace the costly static algorithm. It
is worthwhile to emphasize that the adaptive algorithms will give
more closer results to the static algorithm for denser data, since
the error magnitude is negatively correlated with the item degree and user degree
(for AAF, see Eq. (5) and Eq. (6); for AAS, see Eq. (6)). One may doubt that
although the error caused by the fourth type is small, its accumulative
effects may be considerable and make the approximation farther and
farther to the static algorithm. Fortunately, this will never happen
in real systems because when a new edge adjacent to an item $\alpha$
is added, all the errors of the elements of $\vect{M}$ relevant to
$\alpha$ (i.e., $m_{\alpha \cdot}$ and $m_{\cdot \alpha}$) will be
eliminated by AAS. Therefore, the errors relevant to active items
will not cumulate, while the items rarely attaching new connections
must be less attractive or out of date, which contribute very little
to the recommendations.

\section{Conclusion and Discussion}
Although new data comes every moment, many information filtering
systems still adopt the traditionally algorithmic framework, which
periodically recompute on the whole data. Under the
dynamic environment of data flow, an urgent problem is how to
design effective and efficient adaptive algorithms to give real-time
responses to the change of data. A well-designed algorithm is
supposed to not only provide fast and accurate responses, but also
avoid the cumulative increase of errors, so that the global
recomputation is unnecessary. However, for a general static
recommendation algorithm, there is no guarantee of the existence of
such an adaptive algorithm. In this Letter, being aware of the
reversibility of mass diffusion and heat conduction, we propose a
fast and accurate adaptive algorithm with second-order
approximation, which could follow the static algorithm very tightly
and avoid the cumulative growth of errors.

Although this Letter only presented the adaptive algorithms for mass
diffusion process, the heat conduction process can be handled in an
analogous way. Moreover, the linear integration of mass diffusion
and heat conduction is shown to be able to provide much more
accurate and diverse recommendations than either single algorithm
\cite{Diversity3}. The proposed AAS can be straightforwardly
extended to deal with this linear hybrid case. The present method
can also find its applications for more complicated diffusion-based
recommendation algorithms, for example, the diffusions on multi-rating systems \cite{b.Zhang072,ShangPhysicaA} or hypergraphs involving users, items and tags \cite{Zhang2010,ShangCPL}.

\acknowledgments This work is partially supported by the Swiss
National Science Foundation (Project No. 200020-121848). C.-H.J.
acknowledges the Future and Emerging Technologies programmes of the
European Commission FP7-COSI-ICT (Project QLectives, Grant No.
231200). T.Z. acknowledges the National Natural Science Foundation
of China (Grants Nos. 10635040, 60744003 and 60973069). J.-G.L.
acknowledges the National Science Foundation of China under Grant
No. 10905052.


\begin{thebibliography}{0}

\bibitem{b.Hanani01}
  \Name{Hanani U., Shapira B. \and Shoval P.}
  \REVIEW{User Mod. User-Adap. Interact.}{11}{2001}{203}.

\bibitem{b.Pazzani99}
  \Name{Pazzani M. J.}
  \REVIEW{Artif. Intell. Rev.}{13}{1999}{393}.

\bibitem{b.Goldberg92}
  \Name{Goldberg D., Nichols D., Oki B. M., \and Terry D.}
  \REVIEW{Commun. ACM}{35}{1992}{61}.

\bibitem{b.Resnick97}
  \Name{Resnick P. \and Varian H. R.}
  \REVIEW{Commun. ACM}{40}{1997}{56}.


\bibitem{b.Konstan04}
  \Name{Herlocker J. L., Konstan J. A., Terveen L. G. \and Riedl J. T.}
  \REVIEW{ACM Trans. Inf. Syst.}{22}{2004}{5}.

\bibitem{Adomavicius2005}
  \Name{Adomavicius G. \and Tuzhilin A.}
  \REVIEW{IEEE Trans. Knowl. Data Eng.}{17}{2005}{734}.

\bibitem{Bennett2007}
  \Name{Bennett J. \and Lanning S.}
  \emph{Proc. KDD Cup Workshop} (ACM Press, New York) 2007, pp. 3-6.

\bibitem{Diversity1}
  \Name{Zhou T., Jiang L.-L., Su R.-Q. \and Zhang Y.-C.}
  \REVIEW{Europhys. Lett.}{81}{2008}{58004}.

\bibitem{Diversity2}
  \Name{Zhou T., Su R.-Q., Liu R.-R., Jiang L.-L., Wang B.-H. \and Zhang
  Y.-C.}
  \REVIEW{New J. Phys.}{}{to be published}{}

\bibitem{Diversity3}
  \Name{Zhou T., Kuscsik Z., Liu J.-G., Medo M., Wakeling J. R. \and Zhang
  Y.-C.} arXiv: 0808.2672.

\bibitem{b.Catarina08}
  \Name{Miranda C. \and Jorge A. M.}
  \emph{Proc. the 1st Workshop on Web and Text Intell.} (ICMC/USP, Sao Carlos) 2008, pp. 35-42.

\bibitem{b.Rickard05}
  \Name{Coster R. \and Svensson M.}
  \emph{Proc. the 2005 ACM Sym. on Appl. Comp.} (ACM Press, New York) 2005, pp. 1102-1106.

\bibitem{b.Thomas05}
  \Name{Geogre T. \and Merugu S.}
  \emph{Proc. the 5th IEEE Int. Conf. on Data Mining} (IEEE Computer Society Press, Washington DC) 2005, pp. 625-628.

\bibitem{Suryavanshi05}
  \Name{Suryavanshi B. S., Shiri N. \and Mudur S. P.}
  \emph{Proc. WebKDD Workshop on Tam. Evo., Expan. and Mul.-face. Web Cli.} (ACM Press, New York) 2005, pp. 44-55.

\bibitem{b.Badrul05}
  \Name{Sarwar B., Karypis G., Konstan J. \and Riedl J.}
  \emph{5th Int. Conf. on Com. and Info. Tech.} (IEEE Computer Society Press, Washington DC) 2002, pp. 27-28.

\bibitem{b.Hu08}
  \Name{Wu H., Wang Y.-J. \and Cheng X.}
  \emph{Proc. of the ACM Conf. on Recomm. Syst.} (ACM Press, New York) 2008, pp. 99-106.

\bibitem{b.AlSumait08}
  \Name{AlSumait L., Barbara D. \and Domeniconi C.}
  \emph{Proc. of 8th IEEE Int. Conf. on Data Min.} (IEEE Computer Society Press, Washington DC) 2008, pp. 3-12.

\bibitem{b.Zhang07}
  \Name{Zhang Y.-C., Blattner M. \and Yu Y.-K.}
  \REVIEW{Phys. Rev. Lett.}{99}{2007}{154301}.

\bibitem{b.Zhang072}
  \Name{Zhang Y.-C., Medo M., Ren J., Zhou T., Li T. \and Yang F.}
  \REVIEW{Europhys. Lett.}{80}{2007}{68003}.

\bibitem{b.Zhou07}
  \Name{Zhou T., Ren J., Medo M. \and Zhang Y.-C.}
  \REVIEW{Phys. Rev. E}{76}{2007}{046115}.

\bibitem{Breese1998}
  \Name{Breese J. S., Heckerman D. \and Kadie C.}
  \emph{Proc. 14th Int. Conf. Uncertainty in Artif. Intell.} (Morgan Kaufmann, Madison) 1998, pp. 43-52.

\bibitem{Ren2008}
  \Name{Ren J., Zhou T. \and Zhang Y.-C.}
  \REVIEW{Europhys. Lett.}{82}{2008}{58007}

\bibitem{Shang2009}
  \Name{Shang M.-S., L\"u L., Zhang Y.-C. \and Zhou T.}
  arXiv: 0909.4938.

\bibitem{ShangPhysicaA}
  \Name{Shang M.-S., Jin C.-H., Zhou T. \and Zhang Y.-C.}
  \REVIEW{Physica A}{388}{2009}{4867}.

\bibitem{Zhang2010}
  \Name{Zhang Z.-K., Zhou T. \and Zhang Y.-C.}
  \REVIEW{Physica A}{389}{2010}{179}.

\bibitem{ShangCPL}
  \Name{Shang M.-S. \and Zhang Z.-K.}
  \REVIEW{Chin. Phys. Lett.}{26}{2009}{118903}

\end{thebibliography}
\end{document}